\documentclass[aps,prl,twocolumn,showpacs,floatfix]{revtex4-1}
\usepackage{graphicx,amsfonts,amssymb,amsmath,hyperref}
\usepackage{textcomp}
\usepackage{esint}

\newif\ifhyper
\hypertrue
\ifhyper       
\hypersetup{         
  citecolor = {green},
  colorlinks = {true}, 
  urlcolor = {blue} 
}

\begin{document}

\def\rhoeq{\hat\rho_{\rm eq}}

\newcommand{\marge}[1]{\marginpar{\scriptsize #1}}
\newcommand{\remarque}[1]{\marginpar{\scriptsize Remarque}{\it [#1]}}
\newcommand{\new}[1]{{\bf #1}}
\newlength{\textlarg}
\newcommand{\barre}[1]{%
   \settowidth{\textlarg}{#1}
   #1\hspace{-\textlarg}\rule[0.5ex]{\textlarg}{0.5pt}}
\newcommand{\barred}[1]{%
   \settowidth{\textlarg}{#1}
   \red{#1}\hspace{-\textlarg}\rule[0.5ex]{\textlarg}{0.5pt}}
\newcommand{\barblue}[1]{%
   \settowidth{\textlarg}{#1}
   \blue{#1}\hspace{-\textlarg}\rule[0.5ex]{\textlarg}{0.5pt}}

\def\beq{\begin{equation}}
\def\eeq{\end{equation}}
\def\bleq{\begin{eqnarray}}
\def\eleq{\end{eqnarray}} 
\def\bfig{\begin{figure}}
\def\efig{\end{figure}}
\def\bline{\begin{multline}}
\def\eline{\end{multline}}
\def\bremark{\begin{quotation} \noindent \small }
\def\eremark{\end{quotation}}
\def\llbrace{\left\lbrace}
\def\rrbrace{\right\rbrace}
\def\lbraket{\left[}
\def\rbraket{\right]}
\def\llangle{\left\langle}
\def\rrangle{\right\rangle} 

\newcommand{\Tr}{{\rm Tr}} 
\newcommand{\tr}{{\rm tr}} 
\newcommand{\sgn}{{\rm sgn}} 
\newcommand{\mean}[1]{\langle #1 \rangle}
\newcommand{\commu}[2]{[#1,#2]} 
\newcommand{\bra}[1]{\langle#1|}
\newcommand{\ket}[1]{|#1\rangle}
\newcommand{\braket}[2]{\langle #1|#2\rangle}
\newcommand{\dbraket}[3]{\langle #1|#2|#3\rangle}
\newcommand{\tens}[1]{\overleftrightarrow{#1}}  
\newcommand{\vac}{|{\rm vac}\rangle} 
\def\bravac{\langle{\rm vac}|}
\newcommand{\const}{{\rm const}} 
\newcommand{\atanh}{\,{\rm atanh}}

\newcommand{\ie}{i.e. }
\newcommand{\iet}{i.e.}
\newcommand{\eg}{e.g. }
\newcommand{\cc}{{\rm c.c.}} 
\newcommand{\hc}{{\rm h.c.}} 
\def\etal{{\it et al. }}

\newcommand{\jhatbf}{\hat {\textbf \j}} 
\newcommand{\Jhatbf}{\hat {\textbf \J}} 
\newcommand{\jhat}{\hat {\jmath}} 
\newcommand{\Jhat}{\hat {J}} 
\newcommand{\jbf}{\textbf j}
\newcommand{\Jbf}{\textbf J}

\def\chibf{\boldsymbol{\chi}}
\def\down{\downarrow}
\def\eps{\epsilon}
\def\gam{\gamma} 
\def\phibf{\boldsymbol{\phi}}
\def\varphibf{\boldsymbol{\varphi}}
\def\varphibfs{\boldsymbol{\varphi}_<}
\def\varphibfl{\boldsymbol{\varphi}_>}
\def\varphis{\varphi_{<}}
\def\varphil{\varphi_{>}}
\def\psibf{\boldsymbol{\psi}}
\def\Ome{\Omega}
\def\omeD{{\omega_D}} 
\def\bfOme{\boldsymbol{\Omega}} 
\def\Omebf{\boldsymbol{\Omega}} 
\def\lamb{\lambda}
\def\Lamb{\Lambda}
\def\sig{\sigma}
\def\Sig{\Sigma}
\def\sigp{{\sigma'}} 
\def\bfsig{\boldsymbol{\sigma}} 
\def\sigbf{\boldsymbol{\sigma}} 
\def\bfSig{\boldsymbol{\Sigma}} 
\def\The{\Theta} 
\def\up{\uparrow}

\def\epsk{\epsilon_{\bf k}} 
\def\xik{\xi_{\bf k}} 
\def\txik{\tilde\xi_{\bf k}} 
\def\xip{\xi_{\bf p}} 
\def\xiq{\xi_{\bf q}} 
\def\xikq{\xi_{{\bf k}+{\bf q}}} 
\def\Ek{E_{\bf k}} 
\def\Ep{E_{\bf p}}
\def\Eq{E_{\bf q}}
\def\Heff{\hat H_{\rm eff}}
\def\Hem{\hat H_{\rm em}}
\def\Hint{\hat H_{\rm int}}
\def\Hloc{\hat H_{\rm loc}}
\def\HMF{\hat H_{\rm MF}}
\def\Sem{S_{\rm em}}
\def\SMF{S_{\rm MF}} 
\def\SHF{S_{\rm HF}} 
\def\SRPA{S_{\rm RPA}} 
\def\Sint{S_{\rm int}} 
\def\Sloc{S_{\rm loc}}
\def\TN{T_{\rm N}} 
\def\TNHF{T^{\rm HF}_{\rm N}} 
\def\Zloc{Z_{\rm loc}} 
\def\ZMF{Z_{\rm MF}} 
\def\ZHF{Z_{\rm HF}} 
\def\ZRPA{Z_{\rm RPA}} 
\def\RPA{{\rm RPA}}
\def\loc{{\rm loc}} 
\def\pp{{\rm pp}}
\def\ph{{\rm ph}} 
\def\ch{{\rm ch}}
\def\sp{{\rm sp}} 
\def\qtf{q_{\rm TF}}
\def\epstf{\eps^{}_{\rm TF}} 
\def\epsrpa{\eps^{}_{\rm RPA}} 
\def\chinnzpp{\chi_{nn}^{0}{}\!\!\!''}

\def\half{\frac{1}{2}}
\def\dhalf{\dfrac{1}{2}}
\def\third{\frac{1}{3}} 
\def\quarter{\frac{1}{4}}

\def\qr{{\bf q}\cdot{\bf r}}
\def\wt{\omega t} 

\def\a{{\bf a}}
\def\b{{\bf b}}
\def\e{{\bf e}}
\def\f{{\bf f}}
\def\g{{\bf g}}
\def\h{{\bf h}}
\def\k{{\bf k}}
\def\l{{\bf l}}
\def\m{{\bf m}}
\def\n{{\bf n}} 
\def\p{{\bf p}} 
\def\q{{\bf q}}
\def\r{{\bf r}}
\def\t{{\bf t}}
\def\u{{\bf u}}
\def\v{{\bf v}}
\def\x{{\bf x}}
\def\y{{\bf y}} 
\def\z{{\bf z}} 
\def\A{{\bf A}}
\def\B{{\bf B}}
\def\D{{\bf D}} 
\def\E{{\bf E}} 
\def\F{{\bf F}} 
\def\H{{\bf H}}  
\def\J{{\bf J}}
\def\K{{\bf K}} 

\def\G{{\bf G}}
\def\L{{\bf L}}
\def\M{{\bf M}}  
\def\O{{\bf O}} 
\def\P{{\bf P}} 
\def\Q{{\bf Q}} 
\def\R{{\bf R}}
\def\S{{\bf S}}
\def\epsbf{\boldsymbol{\epsilon}}
\def\mubf{\boldsymbol{\mu}}
\def\nablabf{\boldsymbol{\nabla}}
\def\rhobf{\boldsymbol{\rho}}
\def\sigmabf{\boldsymbol{\sigma}} 
\def\Pibf{\boldsymbol{\Pi}}
\def\pibf{\boldsymbol{\pi}}

\def\para{\parallel}
\def\kpara{{k_\parallel}}
\def\kperp{{k_\perp}} 
\def\kperpp{{k_\perp'}} 
\def\qperp{{q_\perp}} 
\def\tperp{{t_\perp}} 

\def\w{\omega}
\def\wn{\omega_n}
\def\wm{\omega_m}
\def\wnu{\omega_\nu}
\def\wp{\omega_p} 
\def\dmu{{\partial_\mu}}
\def\dnu{{\partial_\nu}}
\def\dl{{\partial_l}}  
\def\dt{\partial_t} 
\def\tdt{\tilde\partial_t}
\def\dk{\partial_k}
\def\tdk{\tilde\partial_k}
\def\dx{\partial_x}
\def\dy{\partial_y} 
\def\dtau{{\partial_\tau}}  
\def\det{{\rm det}} 
\def\Pf{{\rm Pf}}

\def\dsum{\displaystyle \sum}
\def\dint{\displaystyle \int} 
\def\intt{\int_{-\infty}^\infty dt} 
\def\inttp{\int_{-\infty}^\infty dt'} 
\def\intk{\int_{\bf k}} 
\def\intkd{\int \frac{d^dk}{(2\pi)^d}}
\def\intq{\int_{\bf q}} 
\def\intr{\int d^dr}  
\def\dintr{\displaystyle \int d^dr} 
\def\intrp{\int d^dr'}
\def\dinttau{\displaystyle \int_0^\beta d\tau}
\def\dinttaup{\displaystyle \int_0^\beta d\tau'}
\def\inttau{\int_0^\beta d\tau}
\def\inttaup{\int_0^\beta d\tau'}
\def\intx{\int d^{d+1}x} 
\def\inttaur{\int_0^\beta d\tau \int d^dr}
\def\intinf{\int_{-\infty}^\infty}
\def\dinttaur{\displaystyle \int_0^\beta d\tau \int d^dr}
\def\dintinf{\displaystyle \int_{-\infty}^\infty}
\def\intw{\int_{-\infty}^\infty \frac{d\w}{2\pi}}
\def\sumr{\sum_{\bf r}} 

\def\calA{{\cal A}}
\def\calB{{\cal B}} 
\def\calC{{\cal C}} 
\def\dt{\partial_t}
\def\calD{{\cal D}}
\def\calF{{\cal F}} 
\def\calG{{\cal G}}
\def\calH{{\cal H}}
\def\calI{{\cal I}}
\def\calJ{{\cal J}}
\def\calK{{\cal K}}
\def\calL{{\cal L}} 
\def\calN{{\cal N}}
\def\calO{{\cal O}}
\def\calP{{\cal P}}  
\def\calR{{\cal R}} 
\def\calS{{\cal S}}
\def\calT{{\cal T}}
\def\calU{{\cal U}}
\def\calX{{\cal X}} 
\def\calY{{\cal Y}} 
\def\calZ{{\cal Z}} 

\def\calbfB{{\bf \cal B}}
\def\calbfF{{\bf \cal F}}

\def\tT{{\tilde T}}
\def\talpha{{\tilde\alpha}}
\def\tdelta{{\tilde\delta}}
\def\teta{{\tilde\eta}} 
\def\tlamb{{\tilde\lambda}}
\def\tmu{{\tilde\mu}}
\def\tphibf{{\tilde\phibf}}
\def\trho{{\tilde\rho}}
\def\tvarphibf{{\tilde\varphibf}} 
\def\tw{{\tilde\omega}}
\def\twn{{\tilde\omega_n}}

\def\asinh{{\rm asinh}} 
\graphicspath{{./figures/}}

\def\TcMF{T_c^{\rm MF}} 

\title{Simulating frustrated magnetism with spinor Bose gases}

\author{T. Debelhoir}
\author{N. Dupuis}
\affiliation{Laboratoire de Physique Th\'eorique de la Mati\`ere Condens\'ee, UPMC, 
CNRS UMR 7600, Sorbonne Universit\'es, 4 Place Jussieu, 
75252 Paris Cedex 05, France}

\date{March 21, 2016} 

\begin{abstract}
Although there is a broad consensus on the fact that critical behavior in stacked triangular Heisenberg  antiferromagnets --an example of frustrated magnets with competing interactions-- is described by a Landau-Ginzburg-Wilson Hamiltonian with O(3)$\times$O(2) symmetry, the nature of the phase transition in three dimensions is still debated. We show that spin-one Bose gases provide us with a simulator of the O(3)$\times$O(2) model. Using a renormalization-group approach, we argue that the transition is weakly first order and shows pseudoscaling behavior, and give estimates of the pseudocritical exponent $\nu$ in $^{87}$Rb, $^{41}$K and $^7$Li atom gases which can be tested experimentally. 
\end{abstract}
\maketitle

{\it Introduction.} Ultracold dilute atomic gases are ideal laboratories for the realization of (quantum) simulators thus providing an alternative approach to numerical simulations for understanding minimal models of condensed-matter systems~\cite{Bloch08}. This is due to the perfect control and tunability of the interactions in these systems. In this Rapid Communication, we show that the phase transition in three-dimensional stacked triangular Heisenberg  antiferromagnets --an example of frustrated magnets with competing interactions-- can be simulated with spinor Bose gases. This opens up the possibility to solve the long-standing controversy about the nature (second or weakly first order) of phase transitions in these frustrated magnets.

\begin{figure}
\centerline{\includegraphics[width=5cm]{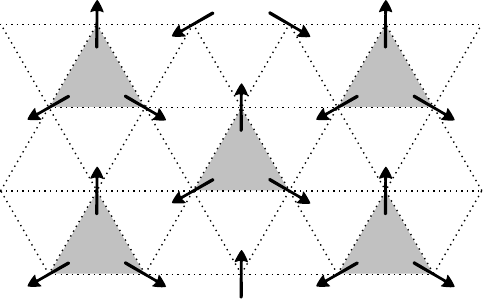}}
\caption{STHA model and ground-state spin configuration.}
\label{fig_STHA} 
\end{figure}

Stacked triangular Heisenberg  antiferromagnets (STHAs) are composed of two-dimensional triangular lattices, with antiferromagnetic coupling between spins, which are piled up in the third direction~\cite{not1}. Because of the frustration due to the antiferromagnetic coupling, the ground state corresponds to a noncollinear spin ordering (with a 120\textdegree~structure, see Fig.~\ref{fig_STHA}). STHAs therefore differ in an essential way from magnets with collinear ordering. In the latter case, the SO(3) spin-rotation invariance is spontaneously broken to SO(2) (corresponding to rotations about the direction of the order parameter) while in STHAs the SO(3) invariance is fully broken, which leads to the existence of three Goldstone modes instead of two when the spin order is collinear. This new symmetry-breaking scheme led to the conjecture that STHAs could be described by a new universality class, different from the O(3)/O(2) universality class of nonfrustrated magnets~\cite{Yosefin85,Kawamura88}. 

Many experiments performed on STHA materials have revealed a continuous phase transition with critical exponents different from those of the O(3) model describing collinear order and different from one compound to the other. These exponents violate scaling relations and the anomalous dimension deduced from the scaling relation $\eta=2-\gamma/\nu$ is negative, which is forbidden by first principles in second-order phase transitions~\cite{Zinn_book}. These results are therefore difficult to reconcile with the standard picture of a second-order phase transition and could indicate that the transition is in fact weakly first order~\cite{Delamotte04}.  

Theoretical studies of frustrated magnets are notoriously hard. Although there is a broad consensus on the fact that the critical behavior in STHAs is described by a Landau-Ginzburg-Wilson Hamiltonian with O(3)$\times$O(2) symmetry, there have been long-standing controversies regarding the nature of the phase transition~\cite{Delamotte04}. On the one hand, in $N$-component frustrated spin models with O($N$)$\times$O(2) symmetry, perturbative renormalization-group (RG) calculations at fixed dimension $d=3$ predict either a second-order phase transition for all $N$ or a window $[N_c^-\simeq 5.7,N_c^+\simeq 6.4]$ of first-order phase transitions~\cite{Pelissetto01a,Calabrese02,Calabrese03,Calabrese04a}. The existence of a focus stable critical point for $N=3$~\cite{not3}, which attracts RG trajectories in a spiral-like approach, leads to unusual crossover regimes and seemingly varying critical exponents that could explain the range of exponents observed experimentally. 
These results have been criticized~\cite{Delamotte10} due to their strong dependence on the resummation parameters but the existence of a stable fixed point 
seems to be corroborated by recent calculations based on the conformal bootstrap program~\cite{Nakayama15,not7}. On the other hand, perturbative RG near $d=4$~\cite{Garel76,Bailin77,Yosefin85,Antonenko95a,Holovatch04,Calabrese04} and the nonperturbative renormalization group (NPRG)~\cite{Tissier00,Tissier03,Delamotte04,Delamotte16} find a first-order phase transition below a critical value $N_c(d)$. The $\eps=4-d$ expansion and the NPRG predict $N_c(d=3)\simeq 6.2$ and $5.1$, respectively, so that the transition for $d=3$ is expected to be first order when $N=3$. 
However, a thorough analysis based on the NPRG has shown that for $N=3$ and $d=3$, even though there is no stable fixed point, the RG flow is very slow in a whole region of the coupling constant space due to an unphysical fixed point with complex coordinates (i.e. a complex solution for the zero of the RG flow)~\cite{Zumbach93,Delamotte04}. This implies the possibility to observe pseudoscaling with effective (nonuniversal) exponents on a large temperature range. The existence of such weakly first-order transitions with pseudoscaling without universality is corroborated by  numerical simulations. (For a summary of experimental, theoretical and numerical issues, see Refs.~\cite{Delamotte04,Delamotte13}.)

In the absence of universality, predicting the pseudocritical exponents of a given material requires to start from a realistic model encoding the lattice structure as well as the microscopic interactions at the lattice scale, a very difficult task in practice. The Hamiltonian of a dilute spin-one Bose gas~\cite{Ohmi98,Ho98,[{For reviews, see }] Kawaguchi12,*Stamper-Kurn13} is similar to the low-energy effective Hamiltonian describing STHAs~\cite{[{A similar analogy between
spin-$\half$ Bose systems and stacked frustrated XY antiferromagnets has been pointed out by }]Ceccarelli15}.
However, in contrast to frustrated magnets, this Hamiltonian is fully determined by a small number of experimentally known parameters, namely the boson mass $M$ and the $s$-wave scattering lengths $a_0$ and $a_2$. This opens up the possibility to 
experimentally test the predictions of the NPRG approach on a quantitative level and therefore 
discriminate between the two theoretical scenarios discussed above for the magnetic transition in STHAs. In particular, the NPRG approach predicts values of the pseudocritical exponent $\nu$ which are significantly different in $^{87}$Rb, $^{41}$K and $^{7}$Li atom gases and can be tested experimentally. 

{\it Frustrated magnets and spin-one Bose gases.} 
The interactions between spins in a STHA are given by the usual lattice Hamiltonian
\beq
H_{\rm lat} = \sum_{\langle i,j\rangle} J_{ij} \S_i \cdot \S_j , 
\label{ham1} 
\eeq
where the $\S_i$'s are three-dimensional vectors of unit length and the sum runs over all pairs of nearest-neighbor sites. The coupling constant $J_{ij}$ equals $J_\para>0$ within the planes and $J_\perp$ in the perpendicular direction~\cite{not1}. In the low-temperature phase where the O(3) invariance of the Hamiltonian~(\ref{ham1}) is spontaneously broken, the order parameter 
\beq
\mean{\S_i} = \a \cos(\Q\cdot\r_i+\theta) + \b \sin(\Q\cdot\r_i+\theta)
\label{op} 
\eeq
corresponding to a (planar) noncollinear ordering can be written in terms of 2 perpendicular vectors $\a$ and $\b$ with equal lengths (Fig.~\ref{fig_STHA}). $\Q=(4\pi/3,0,0)$ is the wavevector of the spin density and we take the distance between nearest neighbors as the unit length. Critical fluctuations are therefore parameterized by 2 three-dimensional vectors $\varphibf_1$ and $\varphibf_2$ such that $\mean{\varphibf_1}=\a$ and $\mean{\varphibf_2}=\b$. The most general form of the effective low-energy Hamiltonian $H[\varphibf_1,\varphibf_2]$ follows from symmetry considerations. The O(3) invariance of $H_{\rm lat}$ implies that $H$ must be invariant in the transformation $\varphibf_1'=R\varphibf_1$ and  $\varphibf_2'=R\varphibf_2$ where $R\in{\rm O}(3)$. In addition, $H$ must be invariant in the O(2) transformation mixing $\varphibf_1$ and $\varphibf_2$: $\varphibf_1'=\cos \alpha\varphibf_1 - \sin\alpha \varphibf_2$ and $\varphibf_2'=\pm(\sin \alpha\varphibf_1 + \cos\alpha \varphibf_2)$. 
This second invariance follows from the arbitrariness of the phase $\theta$ in~(\ref{op}). The Hamiltonian is thus invariant under the symmetry group $G={\rm O}(3)\times {\rm O}(2)$. We can form only two independent ${\rm O}(3)\times {\rm O}(2)$ invariants out of the vectors $\varphibf_1$ and $\varphibf_2$: $\rho=\half(\varphibf_1^2+\varphibf_2^2)$ and $\tau=\quarter( \varphibf_1^2-\varphibf_2^2)^2+(\varphibf_1\cdot\varphibf_2)^2$. To quartic order in the field and lowest order in derivatives, this leads to the effective low-energy Hamiltonian~\cite{Kawamura88}
\beq
H = \int d^3 r \llbrace \half \bigl[(\nablabf \varphibf_1)^2 +(\nablabf \varphibf_2)^2 \bigr] + r \rho + \frac{\lamb_1}{2} \rho^2 + \frac{\lamb_2}{2} \tau \rrbrace ,
\label{ham2}
\eeq
with an ultraviolet momentum cutoff $\Lamb$. 
By choosing $\lamb_2>0$ we ensure that in the ground state $\varphibf_1\perp\varphibf_2$ and $|\varphibf_1|= |\varphibf_2|$ (i.e. $\tau=0$), which corresponds to noncollinear spin ordering (for $\lamb_2<0$, $\varphibf_1$ and $\varphibf_2$ are parallel and the spin ordering is collinear). For $\lamb_2=0$ the Hamiltonian possesses an O(6) symmetry; the transition is second order and belongs to the O(6)/O(5) universality class. 

Let us now consider the Hamiltonian of spin-one bosons~\cite{Ohmi98,Ho98,Kawaguchi12,*Stamper-Kurn13}. Since the total spin is conserved in a binary collision, the interaction Hamiltonian is determined by three potentials $v^{(F)}(\r,\r')$ where $F=0,1,2$ is the total spin of the colliding particles. A classical Hamiltonian describing the critical behavior at the superfluid transition can be obtained by integrating out fluctuations with momenta larger than the inverse of the thermal de Broglie wavelength $\lamb_{\rm dB}=(2\pi/MT)^{1/2}$ (we set $\hbar=k_B=1$). 
Fluctuations with momenta $|\p|\lesssim\lamb^{-1}_{\rm dB}$ behave classically and are described by the (classical) Hamiltonian 
\begin{multline}
H = \beta \int d^3r \biggl\lbrace \sum_m \left[ \frac{|\nablabf\psi_m|^2}{2M}
 - \mu' |\psi_m|^2 \right] \\
+\frac{c_0}{2} \biggl[ \sum_m |\psi_m|^2 \biggr]^2 
+\frac{c_2}{2 }\biggl[ \sum_{m,m'} \psi^*_{m}\F_{m,m'} \psi_{m'} \biggr]^2  \biggr\rbrace  ,
\label{ham3}
\end{multline}
where $\beta=1/T$. The quantum number $m=-1,0,1$ refers to the spin projection on the $z$ axis and $\F\equiv(F^x,F^y,F^z)$ stands for the spin-one matrices. $\mu'$ denotes a renormalized chemical potential. The coupling constants $c_0=(g_0+2g_2)/3$ and $c_2=(g_2-g_0)/3$ are related to the $s$-wave scattering lengths $a_0$ and $a_2$ {\it via} $g_F=4\pi a_F/M$. For symmetry reasons, interactions in the channel $F=1$ are not allowed at low-energy where only $s$-wave scattering is possible~\cite{Kawaguchi12}. 
In principle, $H$ contains terms of arbitrary order, but terms not included in~(\ref{ham3}) are subleading wrt the small parameter $n a_F^3$ and can be ignored ($n$ denotes the density).

Instead of the basis $\lbrace\ket{m}\rbrace$ ($m=-1,0,1$) it is convenient to use the Cartesian basis, defined by $F^\alpha\ket{\alpha}=0$ ($\alpha=x,y,z$), where the field $\psibf=(\psi_x,\psi_y,\psi_z)$ transforms like a vector under spin rotation. The Hamiltonian  
\begin{align}
H ={}& \beta \int d^3r\biggl\lbrace \frac{1}{2M} |\nablabf\psibf|^2 - \mu' |\psibf|^2 \nonumber \\ &
+ \frac{g_2}{2} (\psibf^\dagger\cdot\psibf)^2 
+ \frac{g_0-g_2}{6} |\psibf\cdot\psibf|^2  \biggr\rbrace 
\label{ham4}
\end{align}
is now manifestly invariant under spin inversion and rotations, U(1) (gauge) transformation, and time reversal (complex conjugation) $\Theta$, i.e. $G={\rm O}(3)\times U(1)\times\Theta$. To see the equivalence between Hamiltonians~(\ref{ham2}) and (\ref{ham4}) one writes $\psibf=\sqrt{M/\beta}(\varphibf_1+i \varphibf_2)$ and identifies $r\equiv-2M\mu'$, $\lamb_1\equiv(4M^2/\beta)g_2$, $\lamb_2\equiv(4M^2/3\beta)(g_0-g_2)$ and $\Lamb=\lamb_{\rm dB}^{-1}$. For $\lamb_2>0$ (the case corresponding to noncollinear spin ordering in the STHA), i.e. $g_0>g_2$, the superfluid phase is the so-called ferromagnetic phase~\cite{Kawaguchi12,*Stamper-Kurn13}.

{\it Phase transition in spin-one Bose gases.}
The NPRG approach has been used to study the transition in the ${\rm O}(3)\times{\rm O}(2)$ model~(\ref{ham2})~\cite{Tissier00,Tissier03,Delamotte04,Delamotte16}. Using the equivalence between Hamiltonians~(\ref{ham2}) and (\ref{ham4}) we can use the same approach to make detailed predictions about the transition from the normal phase to the superfluid (ferromagnetic) phase in a spin-one Bose gas. 
All the necessary information is included in the Gibbs free energy (or effective potential) $U(\rho,\tau)$, defined as the Legendre transform of the Helmholtz free energy $-\ln Z$ computed from Hamiltonian~(\ref{ham2}), where $\rho=\half(\phibf_1^2+\phibf_2^2)$ and $\tau=\quarter( \phibf_1^2-\phibf_2^2)^2+(\phibf_1\cdot\phibf_2)^2$ are now functions of the order parameter $\phibf=\mean{\varphibf}$. At the mean-field level, the effective potential $U_{\rm MF}(\rho,\tau) = r \rho + (\lamb_1/2) \rho^2 + (\lamb_2/2)\tau$ is simply given by Hamiltonian~(\ref{ham2}) 
and the transition is second order. 

The Wilsonian RG allows us to include fluctuations in a nontrivial way. In short, one integrates out fluctuations with momenta above a momentum scale $k$ which varies from $\Lambda\sim\lamb_{\rm dB}^{-1}$ down to zero. This defines a momentum-dependent effective potential $U_k$ which is equal to $U_{\rm MF}$ when $k=\Lambda$ and coincides with the (true) Gibbs free energy $U_{k=0}$ when all fluctuations have been integrated out. The NPRG
provides us with an efficient tool to carry out this program~\cite{Berges02,Delamotte12,Kopietz_book}. In practice, we use the LPA' approximation, an improvement of the local potential approximation (LPA) which includes a wavefunction renormalization factor, to solve the exact RG equation satisfied by the effective potential. Furthermore, since $\tau$ vanishes in both the normal and superfluid phases, we use the expansion
\beq
U_k(\rho,\tau) = U_k^{(0)}(\rho) + \tau U_k^{(1)}(\rho) + \frac{\tau^2}{2} U_k^{(2)}(\rho) 
\label{Uexpand}
\eeq
and numerically solve the three coupled RG equations satisfied by the three functions $U_k^{(i)}(\rho)$ ($i=0,1,2$)~\cite{[{A similar expansion has been used previously for a model with U($N$)$\times$U($N$) symmetry: }]Berges97a,*Fukushima11,*Fejos14,*not8}. Note that we make no expansion wrt $\rho$. This allows the description of a first-order transition where a second local minimum may coexist with the minimum at $\rho=0$. 

\begin{figure}
\centerline{\includegraphics[width=5.75cm]{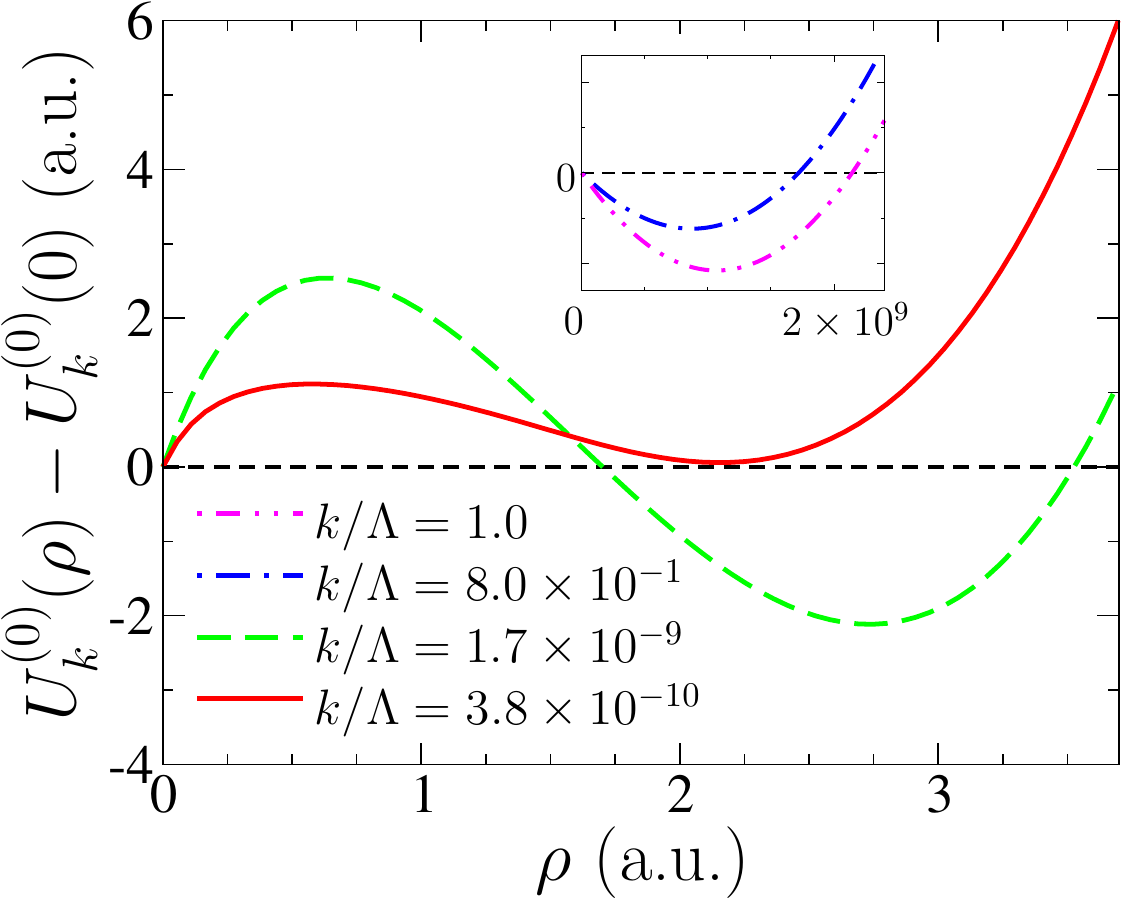}}
\caption{Gibbs free energy $U^{(0)}_k(\rho)$ vs $\rho$ for various values of $k$ and $\mu'=\mu'_c$. The potential exhibits a single minimum at the beginning of the RG flow (see inset) whereas 2 minima coexist for sufficiently small $k$.}
\label{fig_Uk}
\end{figure}

For the numerical solution of the RG equations, we use the known values of $a_0$ and $a_2$ for the Bose gas of interest (e.g. $^{87}$Rb) and choose a typical experimental value for the density $n$~\cite{[{We choose a value of the density corresponding to a recent experiment where the atoms were trapped in a quasi-uniform potential: }]Gaunt13}. We set the temperature equal to its critical value $T_c\simeq(2\pi/M)(n/3\zeta(3/2))^{2/3}$~\cite{not2} and vary the chemical potential $\mu'$ to locate the transition. When lengths are expressed in units of the thermal de Broglie wavelength $\lamb_{\rm dB}$, results are independent of the boson mass $M$. 

Figure~\ref{fig_Uk} shows the $k$-dependence of the effective potential $U_k^{(0)}(\rho)$ at the transition ($\mu'=\mu'_c$) for a $^{87}$Rb atom gas. Initially, for $k=\Lambda$, the system is ordered and $U_k^{(0)}(\rho)$ shows a minimum at a nonzero value $\rho_{0,\Lamb}$. The effect of fluctuations is twofold. Long-range order is suppressed as $k$ decreases (i.e. $\rho_{0,k}$ decreases) and for sufficiently small $k$ a second minimum appears at $\rho=0$.  Both minima become degenerate when $k\to 0$. For $\mu'<\mu'_c$, the minimum at $\rho=0$ is the absolute minimum (normal phase), whereas the nontrivial minimum is the absolute one when $\mu'>\mu'_c$ (superfluid phase). As a consequence the order parameter makes a discontinuous jump at the phase transition, which is therefore (fluctuation-induced) first order. The RG equation $\dk U_k$ is unstable for small $k$ so that it is not possible to determine the effective potential for arbitrary small values of $k$~\cite{not11}. 
Nevertheless, we find that all physical quantities (e.g. the location of the minima of  $U_k^{(0)}(\rho)$ or the correlation length) have nearly converged before the instability occurs~\cite{not6}, and the sole effect of continuing the flow (if it were possible) would be to make the inner part of the potential convex (as required by its definition as a Legendre transform). 

\begin{figure}
\centerline{\includegraphics[width=5.25cm]{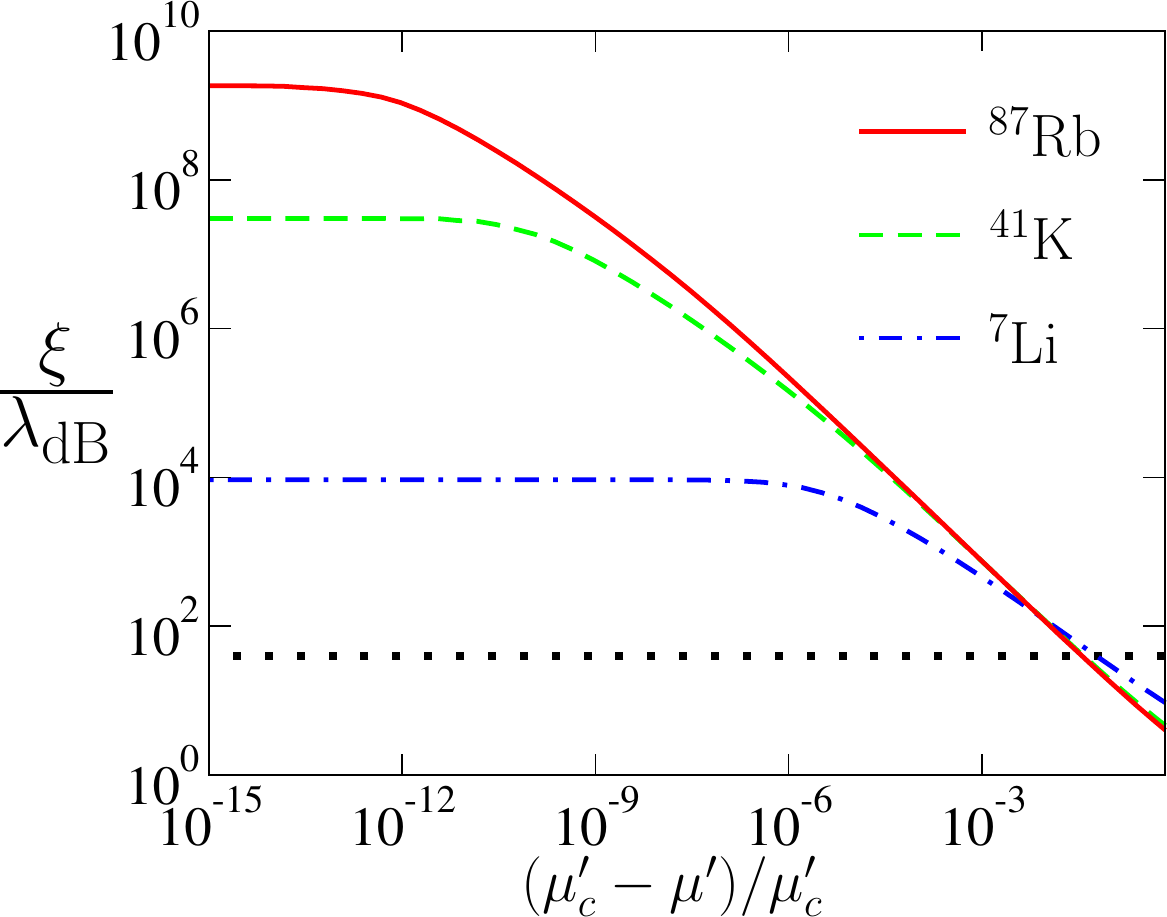}}
\caption{Correlation length $\xi$ vs renormalized chemical potential $\mu'$. The pseudocritical exponent $\nu$ is defined by the slope of the linear dependence of $\ln(\xi/\lamb_{\rm dB})$ vs $\ln(\mu'_c-\mu')/\mu'_c$. The horizontal dotted line shows the size of the system in a typical experiment.}
\label{fig_xi}
\end{figure}

Figure~\ref{fig_xi} shows the correlation length $\xi=(Z_{k=0}/U^{(0)}_{k=0}{}'(0))^{1/2}$ deduced from the one-particle Green function ($Z_k$ denotes the wavefunction renormalization factor and the prime a $\rho$ derivative), obtained from the smallest reachable value of $k$~\cite{not6}, as a function of the renormalized chemical potential. For all atoms considered, $^{87}$Rb, $^{41}$K and $^7$Li, $\xi$ is finite at the transition but several orders of magnitude larger than $\lamb_{\rm dB}$, i.e. much larger than the size $L$ of the system in a typical experiment. This implies that neither the finiteness of $\xi(\mu'_c)$ nor the jump $\Delta n_0$ of the condensate density can be observed experimentally (Table~\ref{table_nu}). 
However, as first pointed out in the context of the magnetic transition in STHAs~\cite{Zumbach93}, the strong increase of the correlation length as the transition is approached allows one to define a (nonuniversal) pseudocritical exponent $\nu$ by $\xi\sim (\mu'_c-\mu')^{-\nu}$ for $\xi\lesssim L$. Note that the same exponent $\nu$ characterizes the increase of the correlation length, i.e. $\xi\sim(T-T_c)^{-\nu}$, if the transition is approached at fixed chemical potential by varying the temperature. Figure~\ref{fig_xi} shows that the regime where pseudoscaling holds is reached as soon as $\xi$ becomes equal to a few de Broglie wavelengths, which suggests that this regime can be observed on a significant temperature range.

Remarkably, the value of $\nu$ differs significantly for $^{87}$Rb, $^{41}$K and $^7$Li and therefore provides us with a possible experimental test of the theory (Table~\ref{table_nu}). The value of $\nu$ varies by less than 2\% if we include only $U_k^{(1)}(\rho)$ in~(\ref{Uexpand}) which shows that the field expansion is nearly converged. Furthermore, higher-order derivative terms not included in the LPA' are expected to be essentially irrelevant for the computation of $\nu$ when, as is the case here, the anomalous dimension $\eta_k=-k\dk\ln Z_k$ is small.
The estimate of $\nu$ is however sensitive to the precise value of the momentum cutoff $\Lamb\sim\lamb_{\rm dB}^{-1}$ (the only parameter which is not precisely known in our theory). For instance, varying $\Lamb$ between $0.25\lamb_{\rm dB}^{-1}$ and $4\lamb_{\rm dB}^{-1}$ one finds $0.55\leq\nu\leq 0.62$ for $^7$Li. 
An improved estimate could be obtained by including quantum fluctuations in the NPRG approach, thus removing the need to introduce a ultraviolet momentum cutoff. Nevertheless, the difference in the value of $\nu$ for $^{87}$Rb, $^{41}$K and $^7$Li is clearly a robust prediction~\cite{not10}.  

\begin{table}
\caption{Correlation length $\xi(\mu'_c)$, condensate-density jump $\Delta n_0$ and pseudocritical exponent $\nu$. The values of $a_0$ and $a_2$ are taken from Ref.~\cite{Stamper-Kurn13}. $a_B$ denotes the Bohr radius and $\lamb_{\rm dB}$ the thermal de Broglie wavelength.}
\label{table_nu} 
\begin{tabular}{cccc}
\hline\hline
 & $^{87}$Rb &$^{41}$K & $^7$Li
\\ \hline 
$a_0/a_B$ & $101.8\pm 0.2$ & $68.5\pm 0.7$ & 23.9
\\ 
$a_2/a_B$ & $100.4\pm 0.1$ & $63.5\pm 0.6$ & 6.8 
\\
$\xi(\mu'_c)/\lamb_{\rm dB}$ & $1.8\times 10^9$ & $3.0\times 10^7$ & $9.1\times 10^3$ 
\\
$\Delta n_0 \lamb^3_{\rm dB}$ & $2.7\times 10^{-9}$ & $1.9\times 10^{-7}$ & $9.3\times 10^{-4}$  
\\ 
$\nu$ & 0.77 & 0.74  & 0.59
\\ \hline \hline
\end{tabular} 
\end{table}

\begin{figure}
\centerline{\includegraphics[width=5.25cm]{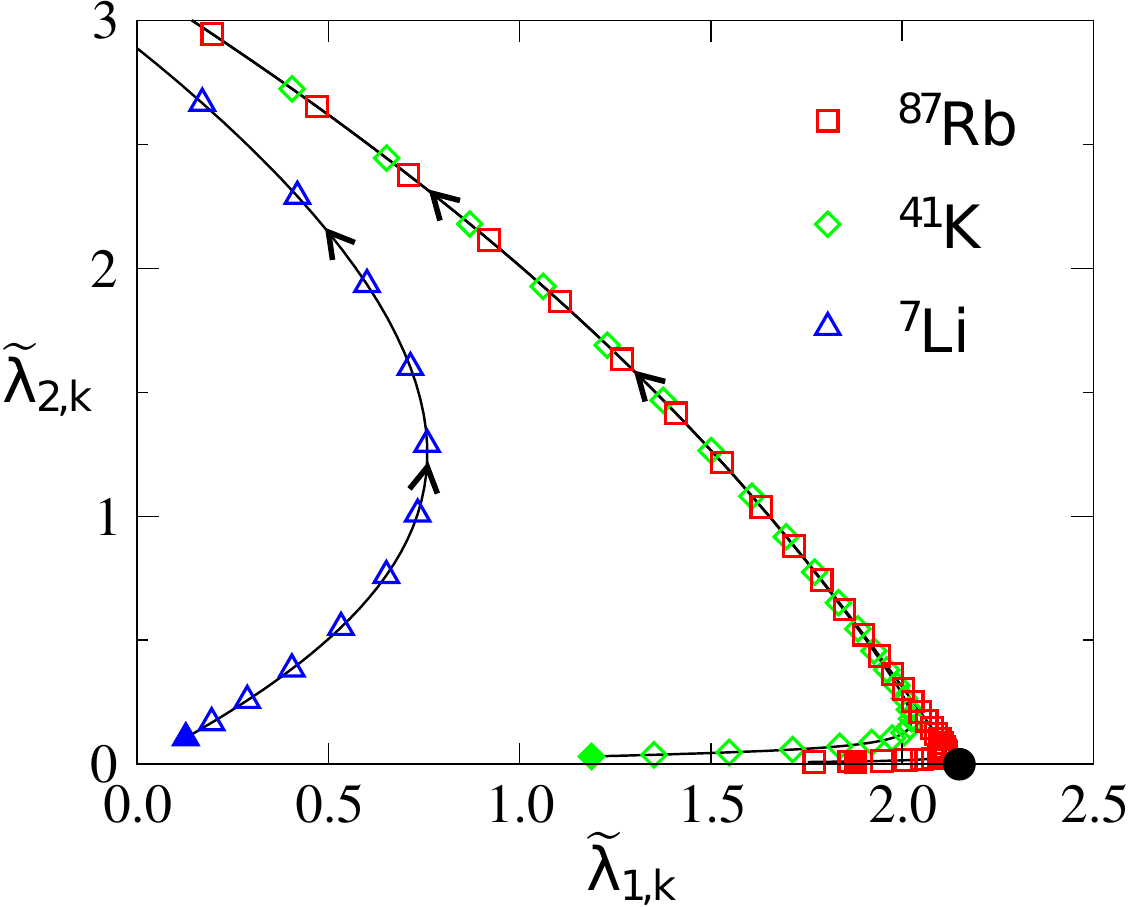}}
\caption{RG flow in the plane $(\tilde\lamb_{1,k},\tilde\lamb_{2,k})$. The solid symbols show the initial conditions of the trajectories. The black dot shows the Wilson-Fisher fixed point of the three-dimensional O(6) model.
}
\label{fig_flow}
\end{figure}

The difference between $^{87}$Rb, $^{41}$K and $^7$Li is illustrated in the RG flow diagram of Fig.~\ref{fig_flow} showing the dimensionless coupling constants $\tilde\lamb_{1,k}=\lamb_1/(Z_k^2k)$ and $\tilde\lamb_{2,k}=\lamb_2/(Z_k^2k)$ as a function of the RG momentum scale $k$. In the case of $^{87}$Rb, for which $\tilde\lamb_{2,\Lamb}$ is very small, the RG trajectory passes near the Wilson-Fisher fixed point of the O(6) model. The RG flow terminates when $k\sim 1/L$ and, for the typical system size $L$ considered, does not leave the region of influence of the O(6) Wilson-Fisher fixed point. It is therefore not a surprise that the pseudocritical exponent $\nu\simeq 0.77$ is not very different from the exponent $\nu_{\rm O(6)}\simeq 0.83$ of the O(6) model. The RG trajectory for $^{41}$K is also strongly attracted by the O(6) Wilson-Fisher fixed point.   
In the case of $^7$Li, pseudoscaling is due to the RG trajectory passing in the vicinity of an unphysical fixed point with complex coordinates and projection (0.690,3.110) in the plane $(\tilde\lamb_{1,k},\tilde\lamb_{2,k})$. In the O($N$)$\times$O(2) model with $N=5$, this unphysical fixed point slows down the RG trajectories and leads to pseudoscaling~\cite{Delamotte04,Zumbach93}. For $N=4$ and $N=3$ no minimum in the velocity of the flow is obtained for $^{87}$Rb, $^{41}$K and $^7$Li, but the RG trajectories in the neighborhood of the unphysical fixed point are nevertheless slow enough for pseudoscaling to be observable. 

{\it Conclusion and experimental discussion.}
We have shown that the O(3)$\times$O(2) model describing the magnetic phase transition in STHAs can be simulated with spin-one Bose gases. The NPRG approach predicts weakly first-order transitions in STHAs and spin-one Bose gases with pseudoscaling without universality. Our predictions can be tested by determining experimentally the correlation length $\xi$ and the pseudocritical exponent $\nu$ using matter-wave interferometry~\cite{Donner07,Navon15}. The value of $\nu$ in $^{87}$Rb and $^{41}$K atom gases, which is close to $\nu_{\rm O(6)}$, is largely a consequence of a crossover phenomenon due to the proximity of the O(6) Wilson-Fisher fixed point, and is independent of the ultimate first-order character of the transition. By contrast, the value $\nu\simeq 0.59$ in $^7$Li is not related in any way to the existence of a nearby critical fixed point: This value is {\it nonuniversal} and depends solely on the scattering lengths $a_0$ and $a_2$. Experimental confirmation of this result would therefore be a 
very strong indication that the weakly-first-order scenario predicted by NPRG is correct.  

One could also consider varying the scattering lengths $a_0$ and $a_2$ by means of a Feshbach resonance. But the external magnetic field, which in general is used to adjust the resonance, would unfortunately suppress the O(3) spin-rotation symmetry. A way out of this difficulty could come from microwave-induced Feshbach resonances as proposed in Ref.~\cite{Papoular10}. Modifying the scattering lengths in $^7$Li would allow a direct confirmation of pseudoscaling, i.e. that the value of $\nu$ changes with $a_0$ and $a_2$. Increasing $a_0$ by a factor of 4 (with $a_2$ fixed) would be sufficient to have $\xi(\mu'_c)<L$ and make the first-order character of the transition observable. 

We would like to thank B. Delamotte, D. Mouhanna and M. Tissier for numerous discussions on the NPRG approach to frustrated magnets, and F. Gerbier and C. Salomon for enlightening discussions on spin-one Bose gases.  



\begin{thebibliography}{44}
\bibitem [{not({\natexlab{a}})}]{not1}%
  \BibitemOpen
  \href@noop {} {} \bibinfo {note} {The sign (ferromagnetic
  or antiferromagnetic) of the coupling $J_\perp$ in the third direction does
  not matter since it does not introduce any frustration.}\BibitemShut {Stop}%
\bibitem [{not({\natexlab{b}})}]{not3}%
  \BibitemOpen
  \href@noop {} {} \bibinfo {note} {A focus fixed point is
  defined by a stability matrix having complex eigenvalues.}\BibitemShut
  {Stop}%
\bibitem [{not({\natexlab{f}})}]{not2}%
  \BibitemOpen
  \href@noop {} {} \bibinfo {note} {Here we use the
  noninteracting value of the transition temperature. For a dilute gas,
  corrections due to interactions are small (of order
  $a_Fn^{1/3}$).}\BibitemShut {Stop}%
\bibitem [{not({\natexlab{g}})}]{not11}%
  \BibitemOpen
  \href@noop {} {} \bibinfo {note} {This instability is due
  to a pole appearing in the propagator at a finite value $k_c$ of the RG
  momentum scale $k$, which prevents to continue the flow for
  $k<k_c$.}\BibitemShut {Stop}%
\bibitem [{not({\natexlab{h}})}]{not6}%
  \BibitemOpen
  \href@noop {} {} \bibinfo {note} {In practice, we
  extrapolate the results to $k=0$ in order to obtain high-precision
  values.}\BibitemShut {Stop}%
\bibitem [{not({\natexlab{i}})}]{not10}%
  \BibitemOpen
  \href@noop {} {} \bibinfo {note} {In Ref.~\cite{Natu11} it
  has been predicted using a mean-field approximation that the (second-order)
  superfluid transition to the ferromagnetic phase becomes a ferromagnetic
  transition without Bose-Einstein condensation (BEC) when $a_0/a_2>5/2$, which
  is the case in the $^7$Li atom gas. Discarding the first-order character of
  the superfluid transition discussed in this Rapid Communication, we expect fluctuations
  beyond mean field to increase the superfluid transition temperature but
  decrease the ferromagnetic transition temperature. Assuming that the shift of
  the BEC temperature is given by $\Delta T_c/T_c\simeq 1.32
  (a_0+5a_2)n^{1/3}$~\cite{Arnold01} and using the mean-field results
  of~\cite{Natu11} for the transition temperature of the ferromagnetic
  transition (without BEC), we find that $^7$Li undergoes a superfluid
  transition rather than a ferromagnetic transition without BEC.}\BibitemShut
  {Stop}%
\end{thebibliography}


%

\end{document}